# Verification of Design Decisions in Communication Protocol by Evaluation of Temporal Logic Formulas

by

Wiktor B. Daszczuk

ICS Research Report 22/98

## Summary

During the project of a communication protocol, many design decisions influence the behavior of the protocol and its correctness. Formal specification and verification of the protocol may prove its correctness. In this paper, an example of a verification of design decision using formal specification in CSM automata and verification in temporal logic is presented.

## 1. Construction of a message protocol

Most protocols (especially low-level protocols) are based on a message-acknowledgment principle. A sender transmits messages, while a receiver transmits acknowledgments. In simplest protocol a message is sent as many times (each time but first after a send timeout occurrence) as long no acknowledgment is received by a sender. In more efficient protocols an acknowledgment may be positive or negative. Negative acknowledgment is sent by a receiver when the message is distorted or a receive timeout occurs. Getting a negative acknowledgment, a sender is informed that the message was distorted or lost quicker than after a timeout.

In alternating bit protocol, the same simple mechanism is used as both positive and negative acknowledgments. Messages and acknowledgments are numbered modulo 2. While sending a message number '0', and waiting for acknowledgment '0', a sender is in a state '0'. Sending message '1' and waiting for acknowledgment '1', a sender is in state '1'.

While a receiver sends an acknowledgment '0' and waits for a message '1', a receiver is in state '0'. Sending acknowledgment '1' and waiting for message '0', a receiver is in state '1'.

Acknowledgments are used as negative in such manner: if a receiver is in state '0', and a receive timeout occurs, or a distorted message arrives, a receiver sends again an acknowledgment '0' (despite of the fact, that a message '0' has been received successfully). The same scheme is used is a sender: if a sender is in state '0', and a send timeout occurs, or a distorted acknowledgment arrives, a sender sends again a message '1'.

## 2. Methods of analysis of the protocol

During the design of the protocol, several decisions must be made. The decisions concern e.g. the size of window, mode of acknowledgments (positive/negative), timeouts, message size etc. These decisions influence the correctness of the protocol as well as its performance.

Performance measurements may be done in real world or using a simulator of the protocol. Both methods allow the designer to measure the throughput of the protocol at various error rates, the cost of transmission of a certain amount of data and similar parameters. But neither measurement nor simulation can prove that the protocol is designed properly, that it has not any logical errors.

The questions of logical correctness can be asked and answered in formal models, such as CSM automata [Mieś92a, Mieś92b]. In this formalism, a designer models the behavior of components as automata, then, he or she may generate the reachability graph of the system using COSMA software [Lach97], and inspect it manually, or alternatively he or she may ask questions formulated in temporal logic and run a TempoRG program [Dasz98] to evaluate them.

In this paper, we will solve the problem if we can use timeouts only in one site (sender or receiver) of the Alternating Bit Protocol (ABP) rather than in both sites. We may treat ABP as an example of more complicated protocol, with reduction of window size to 1.

## 3. Model of the protocol

First we will model the protocol with timeouts in both sites. The automata constituting the model are shown in Fig.1. We define two automata representing *SENDER* and *RECEIVER*. The *SENDER* sends a '0' message and waits for '0' acknowledgment: its a '0' state. If it gets an acknowledgment other than '0', or a send timeout occurs, the *SENDER* sends the '0' message again. If the *SENDER* gets a '0' acknowledgment , it switches its state to '1' and sends a '1' message. The situation in state '1' is symmetric.

The *RECEIVER* waits for a '0' message: it is a '1' state. If it gets a message other than '0', or a receive timeout occurs, the *RECEIVER* sends a '1'acknowledgment (for the previous message '1') again. If the *RECEIVER* gets a '0' message, it switches its state to '0' and sends a '0' acknowledgment. The situation in state '0' is symmetric. The exception is an initial state – the *RECEIVER* does not send '1' acknowledgment until it gets first '0' message form *SENDER*.

To model the medium in which messages and acknowledgments may be lost, two additional automata are designed: *SENDCHANNEL* and *ACKCHANNEL*. The former automaton gets messages '0' and '1', but it may either forward the message to the *RECEIVER* or loose it. The losing of a message is modeled by a signal *slost* external to the model (coming from outsiede of the model). The latter automaton models the medium for acknowledgments, with *rlost* external signal.

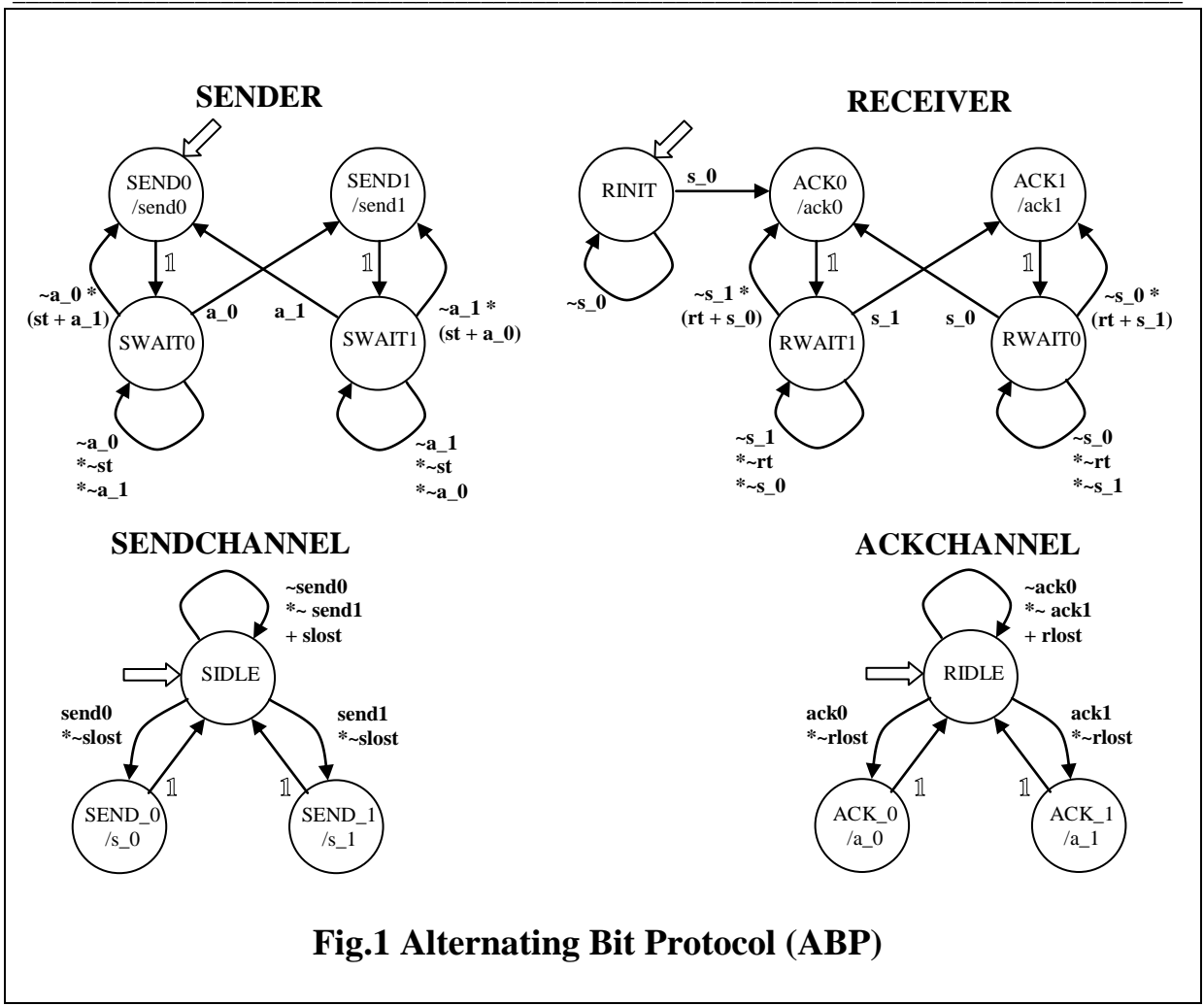

**Fig.1 Alternating Bit Protocol (ABP)**

## 4. Analysis – reachability graph

Fig. 2 presents the shape of the reachability graph of the system presented in Fig. 1. Fig. 2 presents a general view of the reachability graph of the protocol, with four quarters and their side branches shown schematically. Black circles constitute initial path and main loop of the protocol. The main loop bay be divided into four quarters. In every quarter *SENDER* and *RECEIVER* stay in specific states: *BIT:0,ACK:0;  BIT:1,ACK:0;  BIT:1,ACK:1;  BIT:0,ACK:1*.

There are side branches shown schematically as while circles in the figure. In reachability graph side branch is constituted by a set of states, which lead at last to the states of the same quarter of main loop as the side branch was originated form.

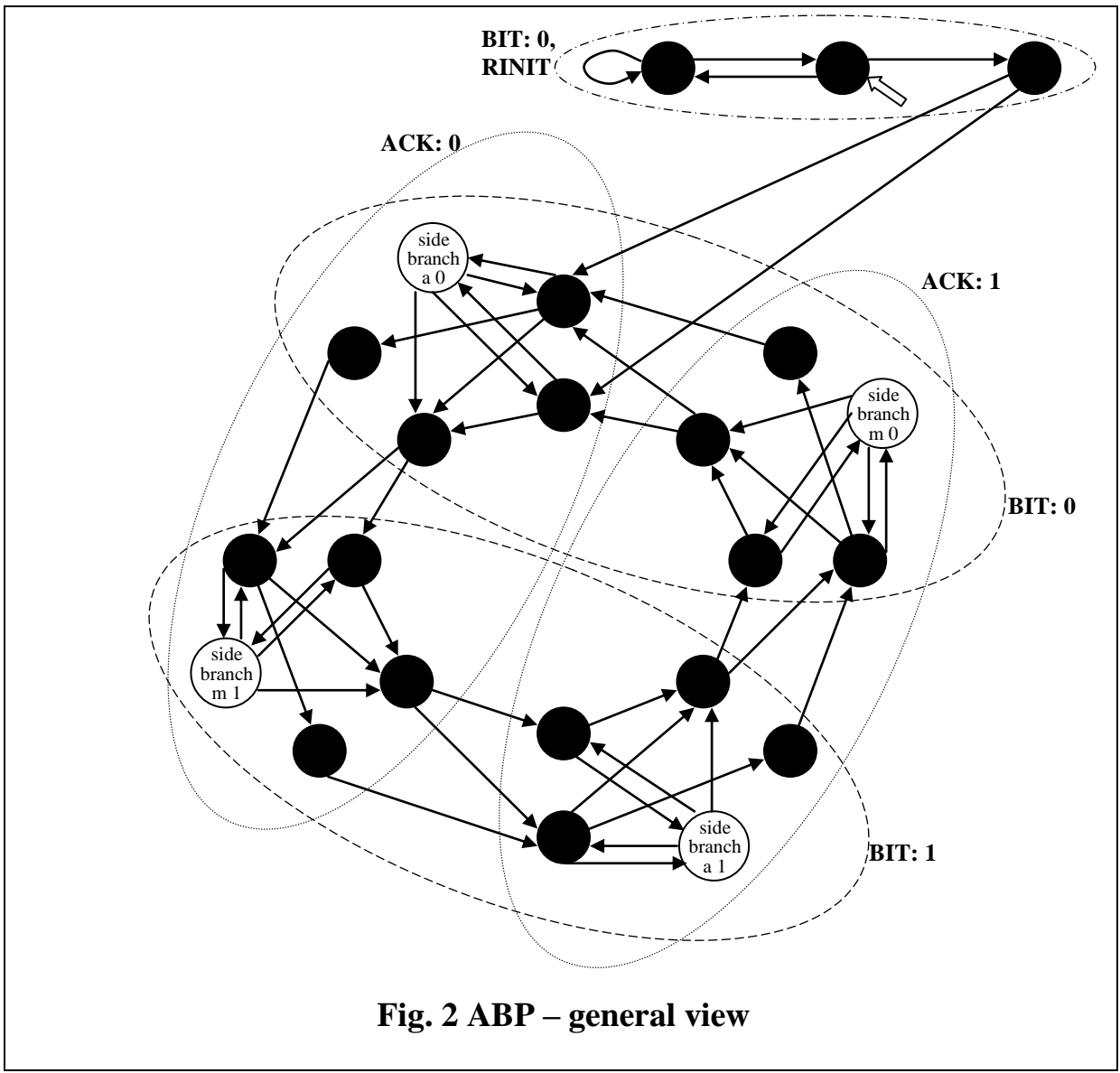

**Fig. 2 ABP – general view**

Fig.3 shows a main loop of the protocol. It represents sequences of states, in which no message or acknowledgment are lost (the situation when a message or an acknowledgment is sent for the second time after a timeout, and it may be lost or not, but the first message/acknowledgment has reached its destination, is not critical and does not lead outside the main loop).

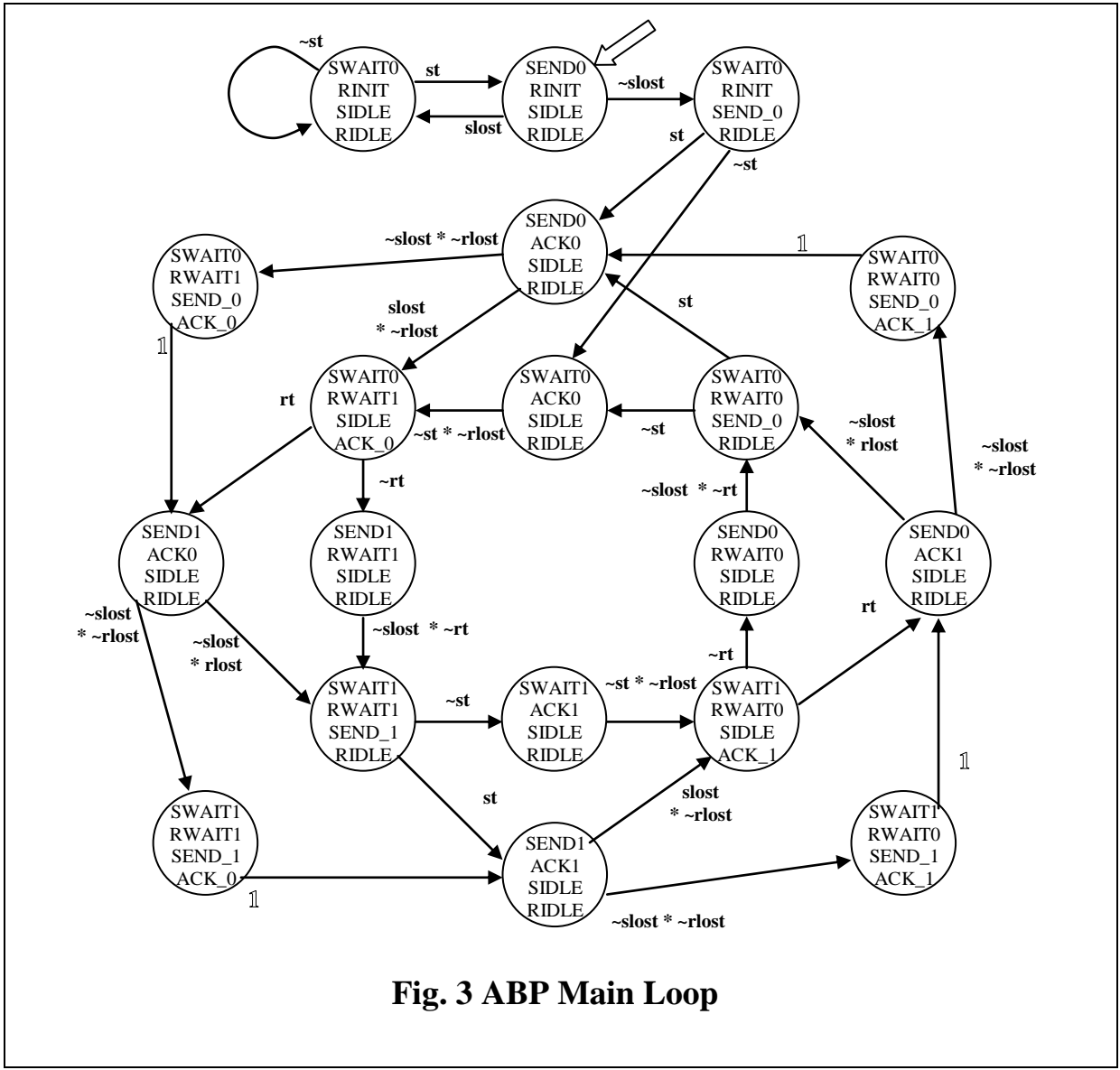

**Fig. 3 ABP Main Loop**

Fig. 4 shows a fragment of the main loop – a first quarter – with a side branch reached when '0' acknowledgment is lost and the protocol requires its retransmission. The protocol may return to the main loop when the '0' acknowledgment is sent for the second time and not lost. Similar side branches are for message '1', acknowledgment '1' and message '0'.

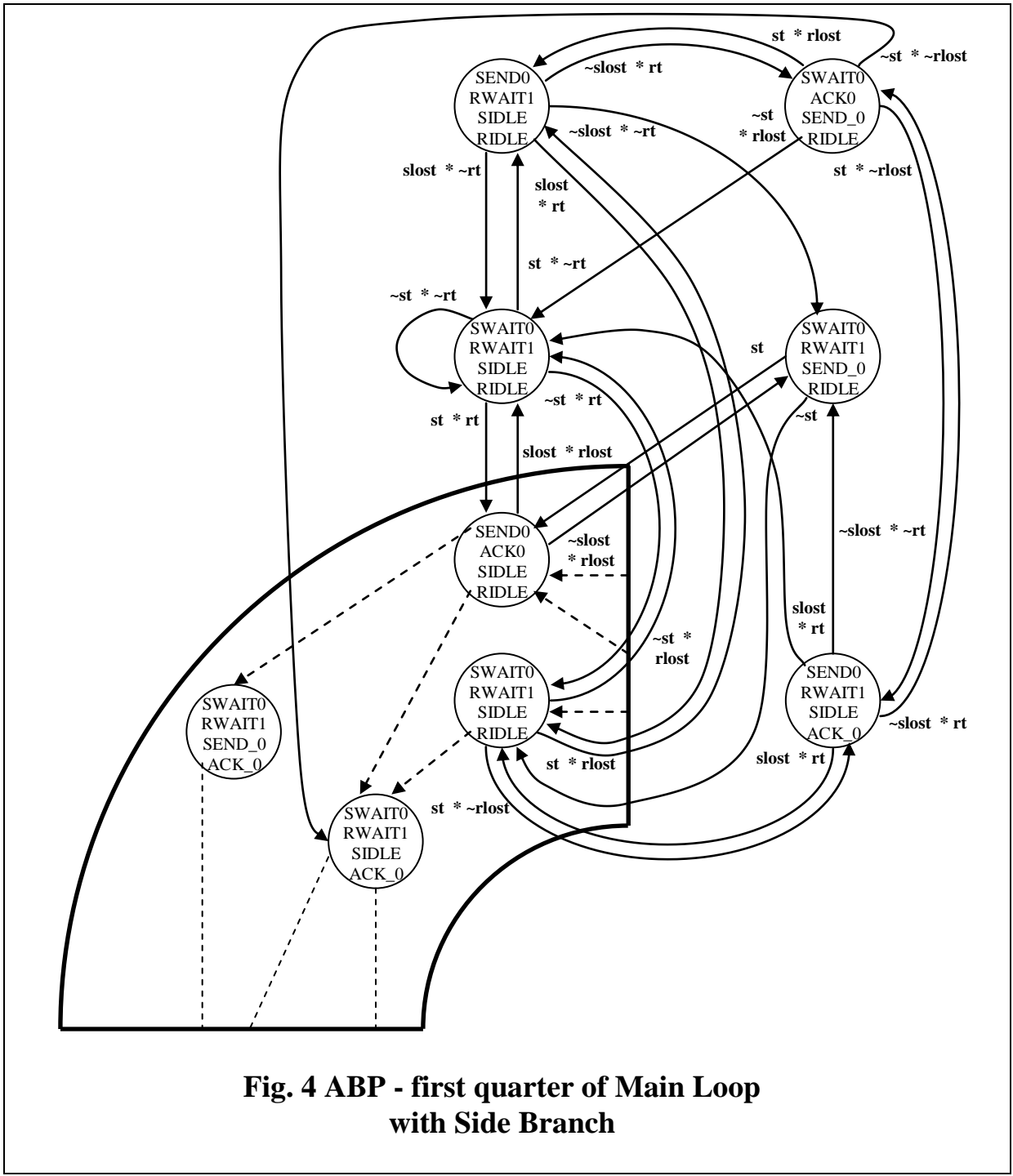

**Fig. 4 ABP - first quarter of Main Loop
with Side Branch**

We may say that the protocol is designed properly, because:
- no deadlock occurs (state with no future),
- no livelock occurs (a dead cycle with no escape and no progress of the protocol),
- The sequence of super-states is proper (the only valid sequence is *BIT:0,RINIT; BIT:0,ACK:0; BIT:1,ACK:0; BIT:1,ACK:1; BIT:0,ACK:1*).

# 5. Formal analysis

Now we will specify temporal formulas that formally express the outlined features of the protocol. Traditional notation is presented first, and then input of the TempoRG program is shown, together with the response of the program. Traditional notation uses a set of operators:

    $\forall$     general quantifier
    $\exists$     specific quantifier (not used in this paper)
    in s     staying in a state s (occurrence of state s)
    $\Diamond \varphi$     formula $\varphi$ will be fulfilled in some state in future (independently on nondeterministic choices)
    $\Box \varphi$     formula $\varphi$ will be held in every state in future (not used in this paper)
    $\circ \varphi$     formula $\varphi$ will be fulfilled in next state
    $\Rightarrow$     implication
    ~     boolean negation
    $\wedge \vee$     boolean product and alternative

After a value of every formula (`--> TRUE` or `--> FALSE`), evaluation time is displayed (hours:minutes:seconds/hundredths) on a computer with Pentium 233 MHz processor and Windows '95 operating system)

Test for deadlock. If there is a path from every state, no deadlock occurs.

$\forall$ s; in s $\Rightarrow$ ($\Diamond$ ~ (in s))
```
A s; in s => (F ! (in s))
 --> TRUE
 Evaluation time is 00:00:00/11
```

Test for liveness. If the protocol is in a gives state (excluding states with *RINIT* state of *RECEIVER*), it eventually reaches this state in future. For the purpose of this question, we will define a set *Aft_INIT* – set of all states that do not have the *RINIT* state of *RECEIVER* in them.

```
[SETS]
Aft_INIT=~{RECEIVER.RINIT}
```

$\forall$ s$\in$Aft_INIT ; in s $\Rightarrow$ ($\circ \Diamond$ in s)
```
A s elof Aft_INIT; in s => (N F in s)
 --> TRUE
 Evaluation time is 00:00:00/22
```

Test for a kind of „liveness" for states including *RINIT* state: they should always lead to the cyclic part of the protocol.

$\forall$ s$\notin$Aft_INIT ; in s $\Rightarrow$ ($\Diamond$ in Aft_INIT)
```
A s elof ~Aft_INIT; in s => (F in Aft_INIT)
 --> TRUE
 Evaluation time is 00:00:00/11
```

Test if every signal is transmitted at last: firstly a test for signal *send0*, which should lead to *s_0*:

send0 $\Rightarrow$ ($\Diamond$ s_0)
```
send0 => (F s_0)
 --> TRUE
 Evaluation time is 00:00:00/05
```

The formula is true, yet this is a naive approach, because the transmission of a *send0* signal through a medium as *s_0* may occur in a next cycle of the main loop.

The formula that better test this sequence of signals is as follows ("no *send1* or *s_1* singnal occurs after *send0*, until *s_0* is sent"):

send0 $\Rightarrow$ ((~(send1 $\vee$ s_1)) U s_0)
```
send0 => ((!(send1 + s_1)) U s_0)
 --> FALSE
 Evaluation time is 00:00:00/00
```

The formula is false ! It is the consequence of naive thinking: not after every *send0* signal its transmission through a medium is required. If a signal *send0*, sent for a second time (third, fourth, ...) after a timeout, is lost but previous *send0* signal reached is destination (the *RECEIVER*) as *s_0*, everything is right. The formula that really test whether every signal reaches the destination is as follows:

$\forall$ s $\in$ {a_1}$\cap${SENDER.SWAIT1}; $\circ$ send0 $\Rightarrow$ ((~(send1 $\vee$ s_1)) U s_0)
```
A s elof {a_1} AND {SENDER.SWAIT1} ; N send0 => (!(send1 + s_1)) U s_0)
 --> TRUE
 Evaluation time is 00:00:00/00
```

Where *{a_1}* denotes a set of all states in reachability graph that generate the signal *a_1*, and *{SENDER.SWAIT1}* denotes a set of all in reachability graph that have the state *SWAIT1* of *SENDER* in them. The formula really test the condition, we ca read it "for all states that send the *send0* signal for the first time, the signal must be transmitted through the medium before the signal *send1* is sent or transmitted through a medium as *s_1*". The condition "send for the first time" is specified by making this state as "next to the previous state". In this case a previous state before a state that issues first *send0* is *SENDER.SWAIT*, and when the *a_1* signal is present, the next state of sender is *SEND0* which generates *send0*. The only exception is the initial state, in which there is no state preceding *SEND0*. For this case we will prepare a separate formula which takes into account the *RINIT* state of *RECEIVER*:

$\forall$ s $\in$ {RECEIVER.RINIT}$\cap${SENDER.SWAIT1}; $\circ$ send0 $\Rightarrow$ ((~(send1 $\vee$ s_1)) U s_0)
```
A s elof {RECEIVER.RINIT} AND {SENDER.SWAIT1} ; N send0 =>          #
 ((!(send1 + s_1)) U s_0)
 --> TRUE
 Evaluation time is 00:00:00/00
```

The next formulas that test analogous conditions are:

$\forall$ s $\in$ {a_0}$\cap${SENDER.SWAIT0}; $\circ$ send1 $\Rightarrow$ ((~(send0 $\vee$ s_0)) U s_1)
```
A s elof {a_0} AND {SENDER.SWAIT0} ; N send1 => ((!(send0 + s_0)) U s_1)
 --> TRUE
 Evaluation time is 00:00:00/00
```

∀ s ∈ {s_1}∩{RECEIVER.RWAIT1}; ○ ack1 ⇒ ((~(ack0 ∨ a_0)) U a_1)
```
A s elof {s_1} AND {RECEIVER.RWAIT1} ; N ack1 => ((!(ack0 + a_0)) U a_1)
 --> TRUE
 Evaluation time is 00:00:00/00
```

∀ s ∈ {s_0}∩{RECEIVER.RWAIT0}; ○ ack0 ⇒ ((~(ack1 ∨ a_1)) U a_0)
```
A s elof {s_0} AND {RECEIVER.RWAIT0} ; N ack0 => ((!(ack1 + a_1)) U a_0)
 --> TRUE
 Evaluation time is 00:00:00/00
```

∀ s ∈ {RECEIVER.RINIT}; ○ ack0 ⇒ ((~(ack1 ∨ a_1)) U a_0)
```
A s elof {RECEIVER.RINIT} ; N ack0 => ((!(ack1 + a_1)) U a_0)
 --> TRUE
 Evaluation time is 00:00:00/00
```

Next formulas test the proper sequence of states in single automata:

send0 ⇒ (◊ send1)
```
send0 => (F send1)
 --> TRUE
 Evaluation time is 00:00:00/06
```

send1 ⇒ (◊ send0)
```
send1 => (F send0)
 --> TRUE
 Evaluation time is 00:00:00/00
```

ack0 ⇒ (◊ ack1)
```
ack0  => (F ack1)
 --> TRUE
 Evaluation time is 00:00:00/05
```

ack1 ⇒ (◊ ack0)
```
ack1  => (F ack0)
 --> TRUE
 Evaluation time is 00:00:00/00
```

and proper sequence of states combined with their input signals:

send0 ⇒ ((~send1 U a_0)
```
send0 => ((!send1) U a_0)
 --> TRUE
 Evaluation time is 00:00:00/06
```

send1 ⇒ ((~send0 U a_1)
```
send1 => ((!send0) U a_1)
 --> TRUE
 Evaluation time is 00:00:00/00
```

ack0 ⇒ ((~ack1 U s_1)
```
ack0 => ((!ack1) U s_1)
 --> TRUE
 Evaluation time is 00:00:00/00
```

ack1 $\Rightarrow$ ((~ack0 U s_0)
```
ack1 => ((!ack0) U s_0)
 --> TRUE
 Evaluation time is 00:00:00/00
```

## 6. Modification 1 – no timeout in *RECEIVER*

Now we can modify the protocol in a specified manner. Let us first withdraw timeouts in *RECEIVER*. The modified *RECEIVER* is shown in Fig. 5a. Because the automaton *RECEIVER* in state *RWAIT0* waits for *s_0* signal (not for timeout), we can stick the states *RINIT* and *RWAIT0* to a new state *RWAIT0*, and set this state as initial (Fig. 5b).

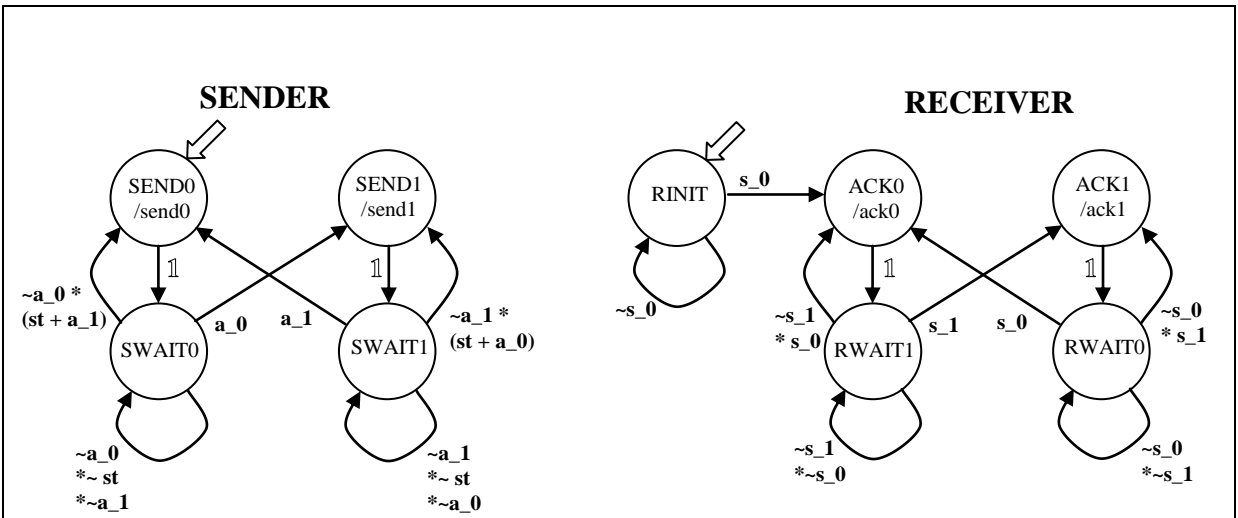

**Fig. 5a ABP – no timeout in RECEIVER**

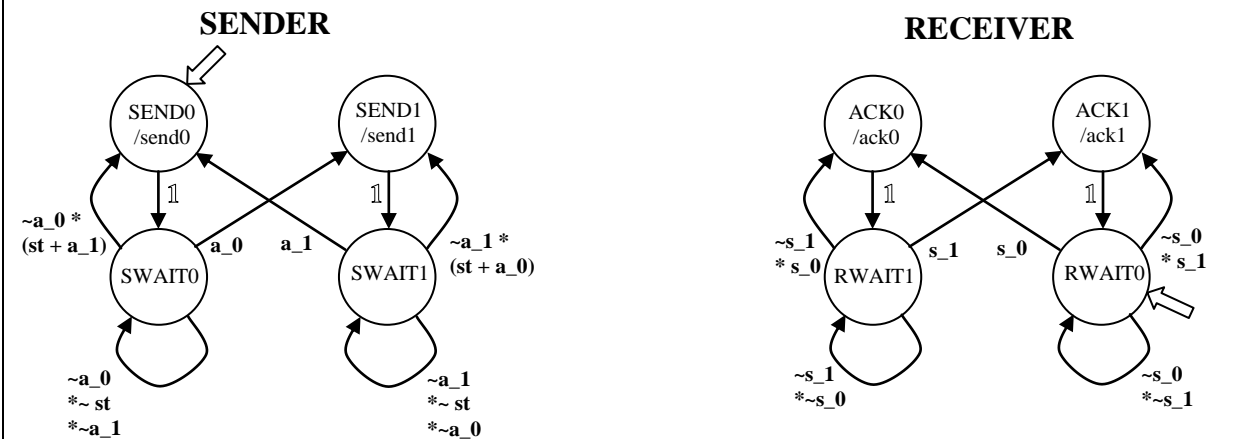

**Fig. 5b ABP – no timeout in RECEIVER simplified version**

The modified protocol fulfils all the correctness conditions outlined in chapter 3. For the protocol without a *RINIT* state in *RECEIVER*, we must modify the conditions of correctness.

$\forall\ s \in \{RECEIVER.RWAIT0\} \cap \{SENDER.SEND0\};\ (\sim(send1 \lor s\_1))\ \mathbf{U}\ s\_0$
```
A s elof {RECEIVER.RWAIT0} AND {SENDER.SEND0} ; (!(send1 + s_1)) U s_0
 --> TRUE
 Evaluation time is 00:00:00/00
```

We must also change the question for liveness, because the *Aft_INIT* set is equal to the full set (no *RINIT* state).

$\forall\ s\ ;\ in\ s \Rightarrow (\circ \Diamond\ in\ s)$
```
A s; in s => (N F in s)
 --> TRUE
 Evaluation time is 00:00:00/11
```

We get positive answer to this question. Thus, we have proven that timeouts may be withdrawn from the protocol in *RECEIVER*.

## 7. Modification 2 – no timeout in *SENDER*

Now we will test if timeouts may be withdrawn in *SENDER*. The modified *SENDER* is shown in Fig. 6a. This solution is erroneous, because the *RECEIVER* waits with timeout for every message other than first. If first message '0' is lost, no negative acknowledgment '1' is sent. This situation leads to a deadlock.

Indeed, a test for deadlock (first correctness condition in chapter 3) is false.

$\forall\ s;\ in\ s \Rightarrow (\Diamond \sim (in\ s))$
```
A s; in s => (F ! (in s))
 --> FALSE
 Evaluation time is 00:00:00/00
```

Thus, we may ask for what states the condition is not held.
```
? s: in s => (F ! (in s))
 --> FULFILLED FOR STATES:
```

The TempoRG program shows the following state:

```
NOT SWAIT0_RINIT_SIDLE_RIDLE
```

This means that a deadlock occurs when *SENDER* is in *SWAIT0* state, *RECEIVER* is in *RINIT* state, and both channels are in their idle states (*SIDLE* and *RIDLE*). Is it exactly the situation, in which first message is lost.

The protocol may be corrected by waiting in *RECEIVER* with timeout for every message, including first. In Fig. 6b, the *RWAIT0* state is set as initial (and state *RINIT* is rejected), which solves the problem. Now the protocol fulfils the correctness conditions outlined in chapter 3. If we ask the same questions as for the protocol in Fig. 5b (because of no *RINIT* state in *RECEIVER*), we will get true for every formula but one.

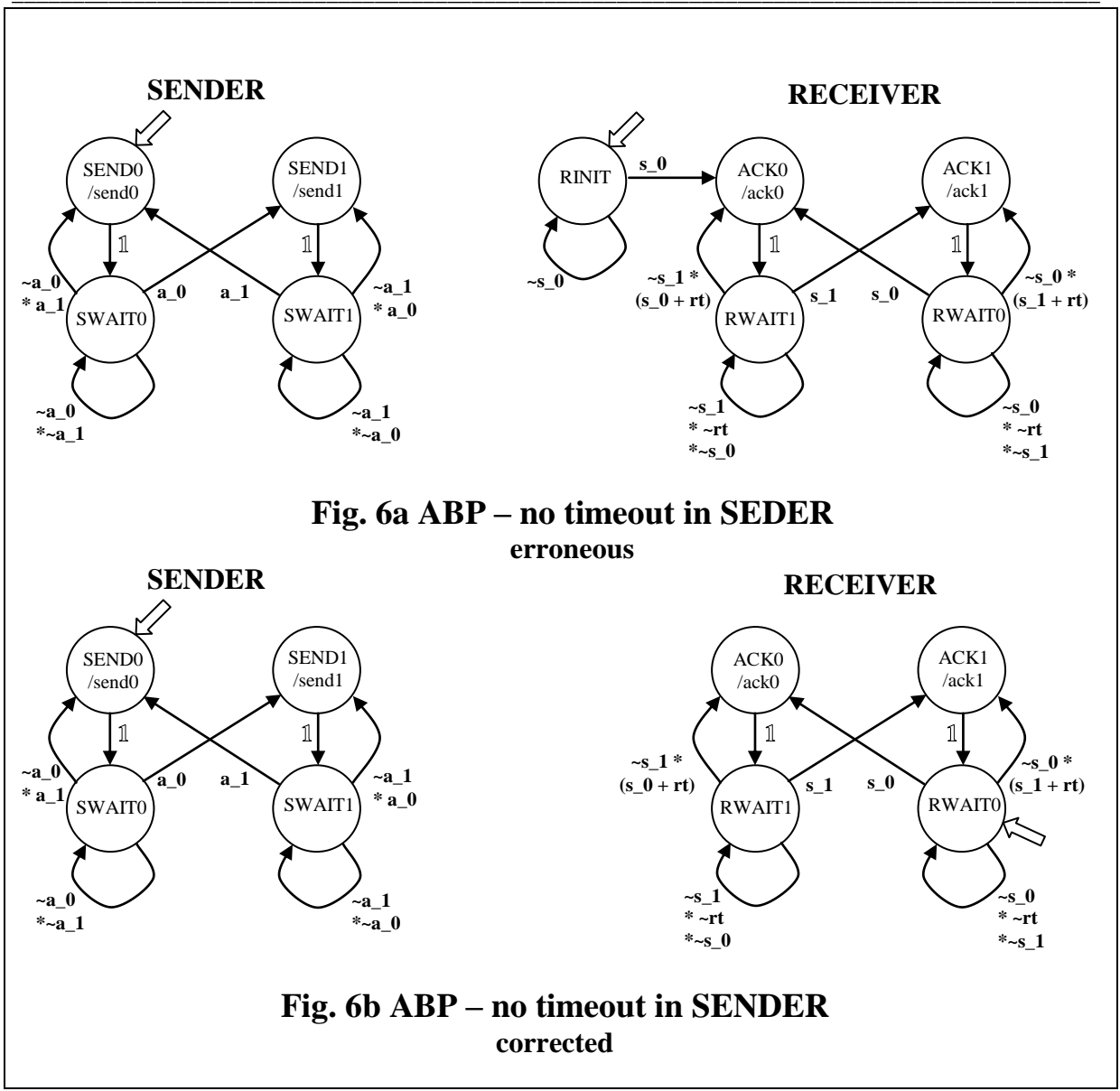

**Fig. 6a ABP – no timeout in SEDER**
erroneous

**Fig. 6b ABP – no timeout in SENDER**
corrected

It is the surprise than we get negative answer to the question for liveness:

$\forall$ s ; in s $\Rightarrow$ ($\circ\Diamond$ in s)
```
A s; in s => (N F in s)
 --> FALSE
 Evaluation time is 00:00:00/06
```

It is because some combinations of states occur only in "initial path" of the protocol and are never reached again, yet the protocol works properly (which is shown by the answers to the rest of questions). Let us ask, which states are in this initial path (no return to them in their future).

```
? s: in s => (N F in s)
 --> FULFILLED FOR STATES:
```

The program lists negative evaluation for:

```
NOT SEND0_RWAIT1_SIDLE_RIDLE
NOT SWAIT0_ACK0_SEND0_RIDLE
```

Now, we may treat these states similarly to the states with *RINIT* state of *RECEIVER* in chapter 3. We will form a set *Aft_INIT* containing all but these two states of reachability graph:

```
[SETS]
Aft_INIT=~{SEND0_RWAIT1_SIDLE_RIDLE,SWAIT0_ACK0_SEND0_RIDLE}
```

$\forall\ s \in Aft\_INIT\ ;\ in\ s \Rightarrow (\circ\Diamond\ in\ s)$
```
A s elof Aft_INIT; in s => (N F in s)
 --> TRUE
 Evaluation time is 00:00:00/05
```

$\forall\ s \notin Aft\_INIT\ ;\ in\ s \Rightarrow (\Diamond\ in\ Aft\_INIT)$
```
A s elof ~Aft_INIT; in s => (F in Aft_INIT)
 --> TRUE
 Evaluation time is 00:00:00/00
```

The answers are true. We have proven that timeouts may be withdrawn from the protocol in *SENDER*.

## 8. Modification 3 – no timeout at all

It is obvious, that a protocol with no timeout (no negative acknowledgment in a case of message or acknowledgment losing) cannot work. But let us test if an analysis of the reachabitily graph confirms this conclusion. The modified automata *SENDER* and *RECEIVER* are shown in Fig. 7a. The protocol does not work – losing of any signal (message or acknowledgment) leads to a deadlock – we have negative answer to the first question. If we withdraw the *RINIT* state from *RECEIVER* – it for sure makes no correction.

$\forall\ s;\ in\ s \Rightarrow (\Diamond \sim (in\ s))$
```
A s; in s => (F ! (in s))
 --> FALSE
 Evaluation time is 00:00:00/00
```

We have proven that in a protocol negative acknowledgments must be sent after a timeout at least in one site: *SENDER* or *RECEIVER*, or in both.

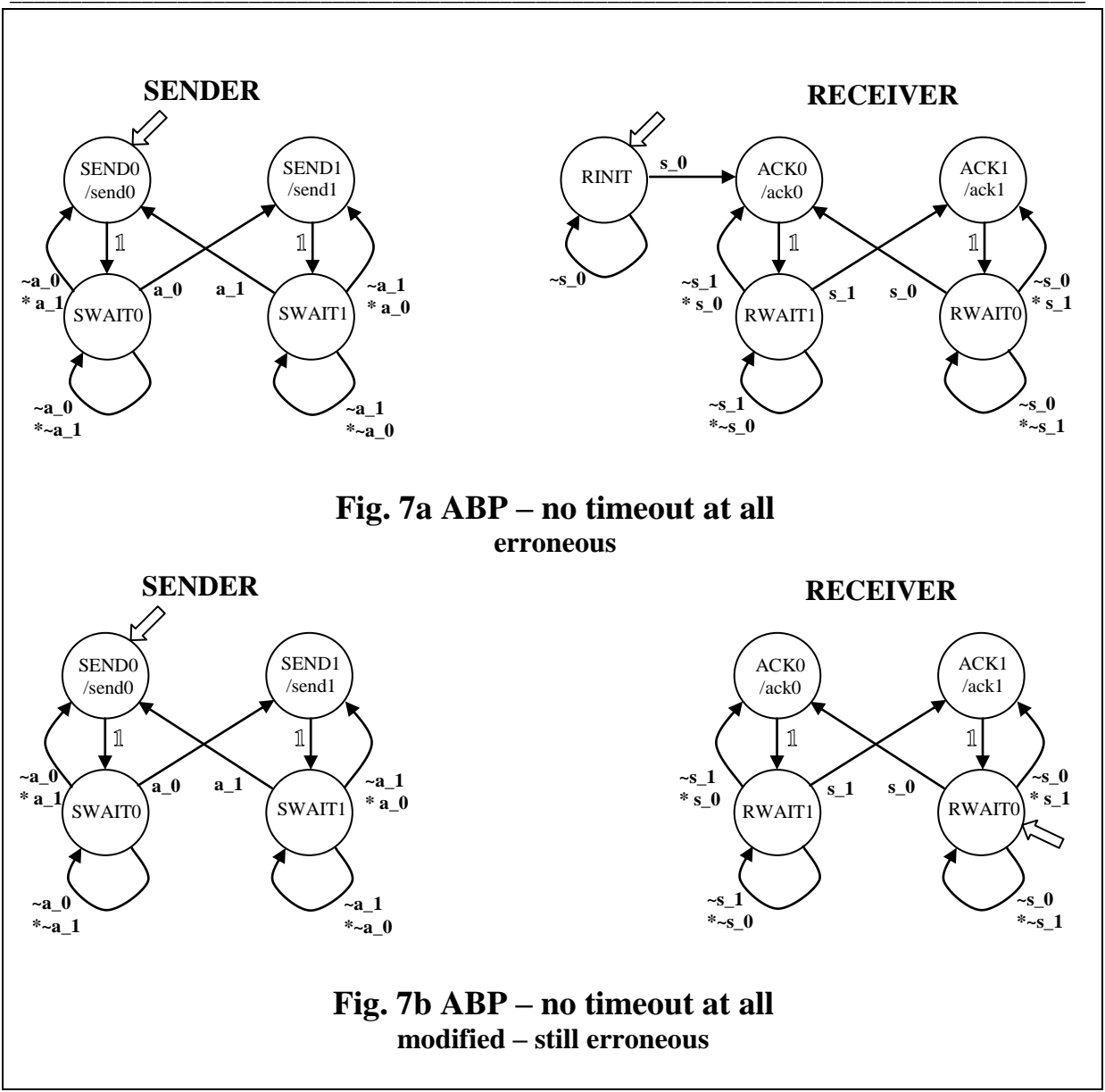

**Fig. 7a ABP – no timeout at all**
erroneous

**Fig. 7b ABP – no timeout at all**
modified – still erroneous

## 9. Conclusions

The design decisions in the project of new protocol are very important. Wrong decisions may lead to improper behavior, or even malfunction. Formal specification and verification of the protocol make it possible to prove that the protocol will work properly.

Specifying the protocol by means of CSM automata and verifying it in temporal logic using TempoRG program is a good way to check design decisions. In this paper, an option of withdrawing timeouts in one of protocol sites is safe. Of course, it may change the performance of the protocol (it may be checked by means of another tool for concurrent system simulation).